\documentclass[conference,a4paper]{IEEEtran}
\usepackage{cite}

\ifCLASSINFOpdf
  \usepackage[pdftex]{graphicx}
   \graphicspath{{C:/Users/Administrator/Desktop/latex/eps}}
  \DeclareGraphicsExtensions{.pdf,.jpeg,.png,.eps}
\else
\fi
\usepackage{CJK}
\usepackage[numbers,sort&compress]{natbib}
\usepackage{cite}
\usepackage{amsmath,amssymb,enumerate}
\usepackage{cite,graphics,graphicx}
\usepackage{graphicx}
\usepackage{subfigure}
\usepackage{listings}
\usepackage{makeidx}
\usepackage{hyperref}
\usepackage{float}
\usepackage{multirow}
\usepackage{amsmath}
\usepackage{amssymb}
\usepackage{xcolor}
\usepackage{epsfig}
\usepackage{graphics}
\usepackage{listings}
\usepackage{algorithm}
\usepackage{algorithmic}
\usepackage{graphicx}
\usepackage{subfigure}
\hypersetup{linkcolor=black, %
            citecolor=black, %
             pdftitle={Title}, %
            pdfauthor={Name}, %
           pdfsubject={Subject}, %
          pdfkeywords={Key words}}
\begin{document}

%
\title { A DCT And SVD based Watermarking Technique To Identify Tag }

\author{\IEEEauthorblockN{Ke Ji{$^{1}$},
Jianbiao Lin{$^{2}$},
Hui Li{$^{3}$},
Ao Wang{$^{4}$},
Tianjing Tang{$^{5}$}
}

\IEEEauthorblockA{
1.2.4.Computer Science and Engineering Department\\ Sichuan University Jinjiang College, Penshan 620860, China\\
3.Colege of Civil Engineering\\ Sichuan University Jinjiang College, Penshan 620860, China\\
}

\IEEEauthorblockA{\IEEEauthorrefmark{1}978864625@qq.com
\IEEEauthorrefmark{2}zeroyuebai@hotmail.com}
}


%


\maketitle

\begin{abstract}
  With the rapid development of the multimedia,the secure of the multimedia is get more concerned. as far as we know , Digital watermarking is an effective way to protect copyright. The watermark must be generally hidden does not affect the quality of the original image. In this paper,a novel way based on discrete cosine transform(DCT) and singular value decomposition(SVD) .In the proposed way,we decomposition the  image into 8*8 blocks, next we use the DCT to get the transformed block,then we choose the diagonal to embed the information, after we do this, we recover the image and then we decomposition the image to 8*8 blocks,we use the SVD way to get the diagonal matrix and embed the information in the matrix. next we   extract the information use both inverse of DCT and SVD, as we all know,after we embed the information seconded time , the information we first information we embed must be changed, we choose a measure way called Peak Signal to Noise Ratio(PSNR) to estimate the similarity of the two image, and set a threshold to ensure whether the information is same or not.  \\

\end{abstract}


\section{Introduction}

Recently, the information hiding technique has developed rapidly in the field of information security and has received significant attention from both industry and academia. It contains two main branches: digital watermarking and steganography. The former is mainly used for copyright protection of electronic product,and it also get rapidly development in recently years.some people use a way to change information sequence instead of changed the embed sequence of the cover[1]. has using both DWT,DCT  and SVD transformation that contributes more robust in comparison with many watermarking algorithms and its use the Zigzag process to embed the information[2].And has proposed a novel way combined Discrete Cosine Transform and Schur Decomposition ,the experiment of this is turn out to be robust.

And there are some research about how to against geometrical attacks.such as [3] has proposed a image watermarking way based on feature point to against geometrical attacks,and the result seems to be successful .The watermarking techniques can be divided into two different classification. One is applied to frequency domain and the other is applied to spatial domain.The spatial domainwatermark are developed early but fragile is their weakness. they can¡¯t against attacks, and the embed information can be easily disfigured, disfigured,or moved. The frequency domain approach has some advantages because most of processing operations can be well characterized, and many good perceptual models are developed in the frequency domain, such as discrete cosine transform(DCT),discrete wavelet transform(DWT), discrete Fourier transform(DFT), Singular value decomposition(SVD),etc. Considered the robustness of the frequency domain, we choose DCT and SVD as two method to embed the same information, and we choose a suitable threshold of PSNR to measure the two image is same or not to judge the product is our product or not.

This paper is organized as follows.Section2 present the theory of Arnold, DCT,SVD,and PSNR.Section 3 describes the watermarking embedding process,In Section 4, the experimental results are shown. The conclusion of this paper is stated in Section 5.

\section{Preliminaries}\label{SEC: Preliminaries}

\subsection{Discrete Cosine Transform(DCT)}\label{SSEC: Discrete Cosine Transform(DCT)}

The main step of the watermarking algorithm based on DCT is let image get DCT first, and then we should choose suitable frequency to embed the information, The last is that we use Inverse Discrete Cosine Transform IDCT to get the the watermarked image. what the  period of the frequency should be choose is a arguable problem, if we choose low frequency  we can get a robust image but we will let the image distortion. if we choose high frequency we will get a opposite result. So consider both robustness and distortion we choose middle frequency to embed the information.

          Figure 2.1 The frequency of the DCT

The formal of DCT and IDCT as followed:

Equation of DCT:

\begin{equation}
F(0,0)=\frac{1}{N}\sum\limits_{x=0}^{N-1}\sum\limits_{y=0}^{N-1}f(x,y)
\end{equation}
\begin{equation}
F(0,v)=\frac{\sqrt 2}{N}\sum\limits_{x=0}^{N-1}\sum\limits_{y=0}^{N-1}f(x,y)\times\cos\frac{(2y+1)v\pi}{2N}
\end{equation}
\begin{equation}
F(u,0)=\frac{\sqrt 2}{N}\sum\limits_{x=0}^{N-1}\sum\limits_{y=0}^{N-1}f(x,y)\times\cos\frac{(2x+1)u\pi}{2N}
\end{equation}
\begin{equation}
F(u,v)=\frac{2}{N}\sum\limits_{x=0}^{N-1}\sum\limits_{y=0}^{N-1}f(x,y)\times\cos\frac{(2x+1)u\pi}{2N} \nonumber \\
\times\cos\frac{(2y+1)v\pi}{2N}
\end{equation}

$F(u,v)$is the element of the transformed matrix,and $f(x,y)$ is the original matrix.

Equation of Inverse DCT:

\begin{eqnarray}
f(x,y)&=&\frac{1}{N}F(0,0)+\frac{\sqrt 2}{N}\sum\limits_{v=0}^{N-1}f(0,v)\cos\frac{(2y+1)v\pi}{2N} \nonumber \\
&+&\frac{\sqrt 2}{N}\sum\limits_{u=0}^{N-1}F(u,0)\cos\frac{(2x+1)u\pi}{2N} \nonumber \\
&+&\frac{2}{N}\sum\limits_{u=0}^{N-1}\sum\limits_{v=0}^{N-1}F(u,v)\cos\frac{(2x+1)u\pi}{2N}\cos\frac{(2y+1)v\pi}{2N} \nonumber
\end{eqnarray}

\subsection{Singular Value Decomposition(SVD)}\label{SSEC: DSingular Value Decomposition(SVD)}

SVD is an effective numerical analysis tool used to analyze matrices. In SVD transformation, a real or complex  $m \times n$ matrix of $A$ of rank $r$ can be decomposed as $A=USV^T$
where $U_{m\times m}$  and $U_{n\times n}$   are unitary  matrix .and   $S_{m\times n}$ is an $m\times n$ rectangular diagonal matrix. Moreover,
\begin{equation}
S_{m\times n}=
\begin{bmatrix}
\Delta_{r\times r} & 0 \\
0 & 0
\end{bmatrix}
\end{equation}

\begin{equation}
\Delta_{r\times r}=diag(\lambda_1,\cdots,\lambda_r)
\end{equation}
$\lambda_1\geq \lambda_2\cdots\geq \lambda_r\leq 0$,$\lambda_{r+1}\geq \lambda_{r+2}\cdots\geq \lambda_n=0$,

We can use a formal as $w=x+¦Á \times i$, x is the original matrix, i is the information ,and ¦Á  is the strength of the information.  and after we embed the information, we should sort the embedded diagonal matrix. We also need a method to  record of this process in order to get the original matrix in later process.

\subsection{Peak Signal to Noise Ratio(PSNR)}\label{SSEC: Peak Signal to Noise Ratio(PSNR)}

PSNR is an engineering term for the ratio between the maximum possible power of corrupting noise that affects the fidelity of its representation.and it is an approximation to human perception of reconstruction quality.Although a higher PSNR generally indicates that the reconstruction of higher quality.

PSNR is most easily defined via the mean squared error(MSE).Given a $m\times n$ image $I$ and $K$, MSE is defined as:

\begin{equation}
MSE=\frac{1}{mn}\sum\limits_{i=0}^{m-1}\sum\limits_{j=0}^{n-1}[I(i,j)-K(i,j)]^2
\end{equation}

The PSNR is defined as:

\begin{equation}
PSNR=10\log_{10}(\frac{MAX_I^2}{MSE})
\end{equation}

Here, $MAX_{I}$is the maximum possible pixel value of the image. When the pixel are represented using 8bits per sample, this is 255.Typical values for the PSNR in lossy image are between 30 and 50 db.

\section{embed and extract process}\label{SEC: embed and extract process}

\subsection{the map of the process}\label{sSEC: the map of the process}

\begin{figure}[!htbp]
\centering
\includegraphics[height=3in]{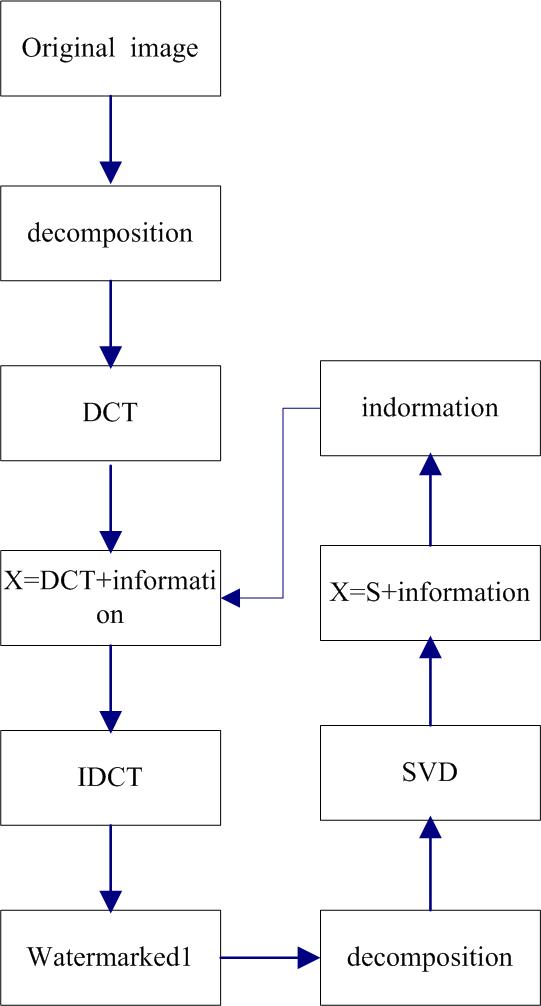}
\caption{THE MAP OF ADD INFORMATION}
\end{figure}

\begin{figure}[!htbp]
\centering
\includegraphics[height=3in]{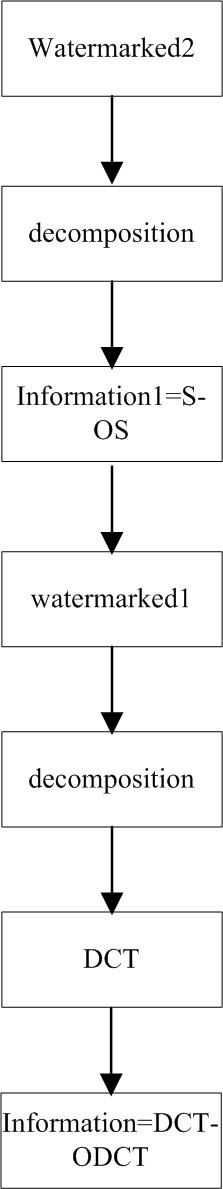}
\caption{ THE MAP OF EXTRACT THE INFORMATION}
\end{figure}

\subsection{The process of add information}\label{SSEC: The process of add information}

 First we need to decomposition the original image to $8\times 8$matrix,then we we use DCT to get a frequency matrix,then we choose a formal as$W=X+\alpha_{i}$; where W is the embed information matrix,X is the original matrix,i is the information. $\alpha$ is the strength of the information ,we choose $\alpha =0.05$.in this paper,Then we need to use IDCT to get the watermarked matrix, and we will represent different $ \alpha $  have different influence on the image in the result.next we will use this watermarked image as a original image, we also need to decomposition the matrix, then we use SVD to get S,and we use formal as \begin{equation} S_{i}=S+\alpha_{i} \end{equation} $ \alpha $ is the strength of the information, i is the information and ${S_{i}}$ is the final watermarked image.

\subsection{The process of extract information}\label{SSEC: The process of extract information}

we need to decomposition the final watermarked to $8\times 8$matrix, then we use a formal \begin{equation}i=(\frac{S_{I}-S}{\alpha}) \end{equation} to get then first information, then we get the image that only have the DCT method information, so we get the image decomposition first, then we use DCT to make the matrix to become a frequency domain, we use the formal \begin{equation} i=(\frac{S-OS}{\alpha}) \end{equation} S is the watermarked transform matrix, and S is the original matrix.

\section{experiment}\label{SEC: experiment}

\subsection{original image}\label{SSEC: original image}

\begin{figure}[!htbp]
\centering
\includegraphics[height=1.6in]{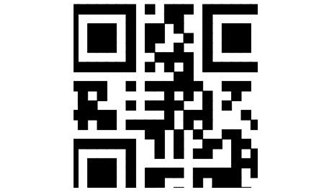}
\caption{watermark image}
\end{figure}

\begin{figure}[!htbp]
\centering
\includegraphics[height=1.5in]{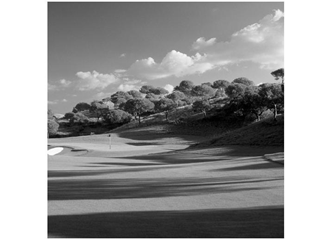}
\caption{original image}
\end{figure}

\subsection{Different values in PSNR  that we choose }\label{SSEC: Different values in PSNR  that we choose  }

$\alpha = 0.05$

\begin{figure}[!htbp]
\centering
\includegraphics[height=1.5in]{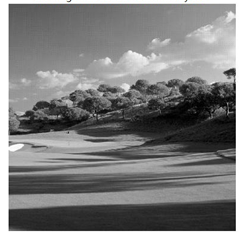}
\caption{the image that added watermark by svd}
\end{figure}

\begin{figure}[!htbp]
\centering
\includegraphics[height=1.5in]{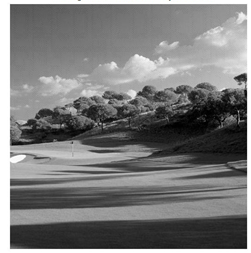}
\caption{The image that added watermark by svd and dct}
\end{figure}

$\alpha = 0.03$

\begin{figure}[!htbp]
\centering
\includegraphics[height=1.5in]{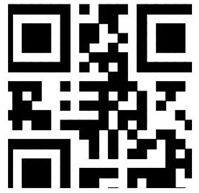}
\caption{The extracted watermark by svd}
\end{figure}

\begin{figure}[!htbp]
\centering
\includegraphics[height=1.5in]{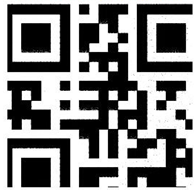}
\caption{The extracted watermark by dct}
\end{figure}

The two watermark of the PSNR=52.4885

 ~\\

\subsection{different noisy in image }\label{SSEC: different noisy in image }

\subsubsection{add gauusian noisy }\label{SSSEC: add gauusian noisy }  \textrm{}

\begin{figure}[!htbp]
\centering
\includegraphics[height=1.5in]{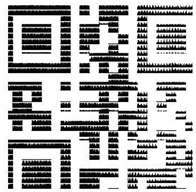}
\caption{The extracted watermark by svd}
\end{figure}

\begin{figure}[!htbp]
\centering
\includegraphics[height=1.5in]{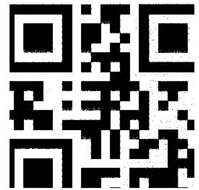}
\caption{The extracted watermark by dct}
\end{figure}

The two watermark of the PSNR=35.8880

\subsubsection{add salt and pepper noisy }\label{SSSEC: add salt and pepper noisy }  \textrm{}

\begin{figure}[!htbp]
\centering
\includegraphics[height=1.5in]{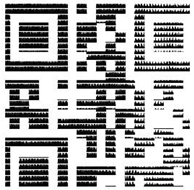}
\caption{The extracted watermark by svd}
\end{figure}

\begin{figure}[!htbp]
\centering
\includegraphics[height=1.5in]{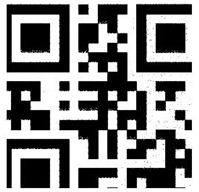}
\caption{The extracted watermark by dct}
\end{figure}

The two watermark of the PSNR=35.8829

\subsubsection{add some random noisy }\label{SSSEC: add some random noisy}  \textrm{}

~\\

\begin{figure}[!htbp]
\centering
\includegraphics[height=1.5in]{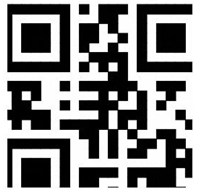}
\caption{The extracted watermark by svd}
\end{figure}

\begin{figure}[!htbp]
\centering
\includegraphics[height=1.5in]{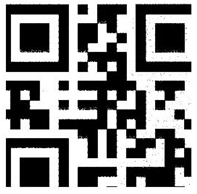}
\caption{The extracted watermark by dct}
\end{figure}

The two watermark of the PSNR=49.2984

\section{Conclusion}\label{SEC: Conclusion}

In this paper, we choose a robust way to identify tag, It combine both DCT and SVD to embed the same information, so if we get attack some noisy in the watermarked image, the result proved to be succeed, but we can¡¯t get a good way to deal with the edge .but there is no doubt that this is a good way to identify tag.It get a robust and invisible watermarked image.

\section{ Acknowledgement}\label{SEC:  Acknowledgement}

The research subject was supported by  the department of Computer Science$\&$Engineering Jinjiang College, Sichuan University.  Thanks for Prof.Bingfa Lee¡¯s suggestions and guidance.




%
\fontsize{8pt}{\baselineskip}\selectfont

\citestyle{IEEEtran}
\bibliographystyle{IEEEtran}


%

\end{document}